\documentclass[10pt,a4paper,fleqn]{article}
\usepackage{amsmath}
\usepackage{amsfonts}
\usepackage{amssymb}
\usepackage{xcolor}
\usepackage[T1]{fontenc}
\usepackage{graphicx}
\usepackage{inputenc}
\usepackage{xurl}
\usepackage{booktabs}

\graphicspath{{./img/}}

\renewcommand{\vec}[1]{\ensuremath{\boldsymbol{#1}}}
\DeclareMathOperator{\var}{Var}
\DeclareMathOperator{\expected}{E}
\DeclareMathOperator{\snr}{SNR}
\DeclareMathOperator{\tpr}{TPR}
\DeclareMathOperator{\fpr}{FPR}
\DeclareMathOperator{\rmse}{RMSE}
\DeclareMathOperator{\rmsle}{RMSLE}

\author{Alessandro Schaer\thanks{Magnes AG, Hardturmstrasse 253, 8005 Zurich, Switzerland.
AS: \texttt{https://orcid.org/0000-0001-9865-9185},
GC: \texttt{https://orcid.org/0009-0001-0728-5286},
HM: \texttt{https://orcid.org/0009-0000-9231-0919}}
\and Henrik Maurenbrecher\footnotemark[1]
\and George Chatzipirpiridis\footnotemark[1]
\and Hamdi Torun\thanks{School of Engineering, Physics and Mathematics, Northumbria University, Newcastle upon Tyne, NE1 8ST, UK. Correspondence: \texttt{hamdi.torun@northumbria.ac.uk}, \texttt{https://orcid.org/0000-0002-7882-286X}.}
}

\title{On the Removal of Artifacts of Known-Shape from Noisy Signals}

\begin{document}
	\maketitle

  \textit{Abstract} --- {\bf A general method for the estimation and removal of quasi-periodic,
  artifact-like disturbances from single channel measurements is presented.
  The method is based on a wavelet template and data-driven template extraction from single channel,
  noisy signals.
  The method is tested on an example application in modern neurology.
  The method is compared to an autoencoder, trained and deployed under idealized conditions, thus
  acting as reference system.
  It is found that the proposed method
  yields signal estimates with median
  root mean squared error improvement
  of $\boldsymbol{33\%}$ compared to the baseline,
  which is $\boldsymbol{4\%}$ more than the autoencoder,
  while relying on fewer assumptions and parameters,
  and without the need for any training data.
  }


  \section{Introduction}\label{sec:introduction}

In all practical cases, measurements are an imperfect representation of what one is really interested in.
Generally, if one is interested in some signal $x$, only some distorted version of it $z = g(x)$ is obtainable in practice.
Here $z$ represents the actual measurement, which is some mapping
$g(\cdot)$ applied to the true state $x$.
Several sources of uncertainty can corrupt any measurement that is performed in real-life, one of the simplest being additive noise, in which case one has $z = g(x) = x + n$, where $n$ is the noise.
The sources of uncertainty may or may not be silenced at the time of measurement, and when they cannot be removed
before measuring there may be the need for separating them from the signal of interest using signal-processing
techniques prior to performing any further analysis on the cleaned signal.
Mathematically, this corresponds to applying some mapping $h(\cdot)$ to the
measurement $z$ to obtain an estimate of $x$, which we denote as $\hat{x}$, or $\hat{x} = h(z)$.
Linear filtering (band-pass/stop) techniques are one example of such mappings $h(\cdot)$.
These filters can be used when the disturbances and
signal of interest are spectrally separated, for example the removal of Mains-Hum in signals with
spectral content far away from the mains frequency (typically 50\,Hz or 60\,Hz) using a notch filter.
Unfortunately, the signal and disturbance often times overlap spectrally, hence additional, more advanced
techniques for disturbance rejection are required.

Several techniques have been developed, tackling the blind-source separation problem, with some
techniques being better suited for specific types of signals than others.
An overview for physiological signals is given in~\cite{sweeney2012artifact}.
These methods enable the identification and separation of the ``source of truth'' from background noise and disturbances, which can then processed further.
Many of them, such as independent component analysis (ICA), rely on having multiple
measurements at hand, which is not always a satisfiable constraint.

For the problem of artifact identification and removal, template matching techniques have long been known,
in particular in computer vision~\cite{brunelli1997template}.
In the analysis of local field potentials (LFP) of the brain, template removal for cleaning LFP signals from
heart artifacts has been used~\cite{neumann2021sensitivity,chen2021removal,hammer2022artifact},
though with somewhat unclear implementations and performance analyses.
Furthermore, it seems that templates are subtracted without any considerations in terms of local-matching --
the template is usually extracted as an averaged kernel, which is subtracted from each identified artifact, without
considerations w.r.t. the specific artifact instance amplitude.

Recently, autoencoders~\cite{kramer1992autoassociative} and other deep-learning models have raised to prominence,
arguably becoming the \textit{de facto} gold standard
for denoising, artifact removal, and fault detection~\cite{xiong2015denoising, zhou2017anomaly,almazrouei2019using,cheng2021improved,saba2021unsupervised,hossain2022deep}.
Machine-learning techniques, such as autoencoders require good training data as their performance degrades due to
inadequate generalizations
Proper care and infrastructure is needed for their application.
In the case where ground-truth data is difficult (or even impossible) to gather, synthetic data can be used for
training, but this only postpones the verification of the performance in real-world deployments.
Should the synthetic data be insufficiently representative of the real-world conditions, there is no easy and fast
way of fixing a model, as the bad-data is what defined the model in the first place.
Hence, more flexible models can be better alternatives to autoencoders in some scenarios.

An interesting problem from a signal-analysis perspective thus remains the removal of artifacts from noisy
signals, when only one measurement is available, the disturbance spectrum and the signal spectrum overlap
significantly, and no reference measurement for the disturbance is available.
In these cases, methods such as ICA are not applicable, reference correlation is not an option, and multiple
sources of distortion have to be coped with simultaneously.
We propose a method to tackle this problem for cases in which a signal is contaminated by both  noise
and strong artifacts based on data-driven template matching.
Our contribution is threefold:
\begin{enumerate}
    \item We illustrate a method for the removal of artifacts of known shape from noisy signals;
    \item We provide open-source Python implementations of the methods described herein including a signal generation package and an artifact removal package; and
    \item We showcase the performance of the proposed method on a sample use-case including a comparison to a machine-learning approach placed within an idealized context -- specifically a common problem in modern neurology, using synthetic data.
\end{enumerate}

To aid the reciprocal understanding, we will be using the following notation and terms throughout this manuscript.
We denote the true signal as $x$, noise with $n$, and disturbance (artifacts) with $d$.
The measurement is denoted by $z$ which is a function of signal, noise, and disturbance $z = g(x,n,d)$.
The goal is to manipulate the measurement $z$ in order to obtain an accurate estimate $\hat{x}$ of the underlying
signal $x$ using some mapping $h$: $\hat{x} = h(z) = h(g(x,n,d))$.
  \section{Methods}\label{sec:methods}

The whole analysis has been implemented in Python 3, leveraging the numerical,
scientific, machine-learning and visualization libraries Numpy, SciPy, SciKit Learn, and Matplotlib~\cite{harris2020array,virtanen2020scipy,pedregosa2011sklearn,hunter2007matplotlib}.
The signal generation and artifact removal codebases are released as open-source
code.

\subsection{Formalized Problem Statement}
We are interested in estimating the damping of some disturbance $d[k]$
affecting the sampled signal $x[k]$ for a set of
measurement points $0 \leq k < N-1$, which is additionally corrupted by noise $n[k]$.
We focus on the case where $g(\cdot)$ is a linear mixing function.
In other terms, we want to reduce the effects of $d[k]$ in the measurement
\begin{equation}
    z[k] = x[k] + \gamma \cdot d[k] + \sigma \cdot n[k], \quad \gamma,\sigma > 0. \label{eq:measurement}
\end{equation}
It is assumed that:
only $z[k]$ for $0 \leq k < N-1$ is available as a measurement;
$x[k]$ and $d[k]$ have (partially) overlapping spectra;
the disturbance $d[k]$ manifests as a recurring ``pulse'' of known shape $\psi[k]$;
and an estimate for the upper-bound of the frequency $f_{d\lim{}}$ of the disturbance pulses is known.
This last assumption can also be interpreted as knowing the minimum distance in time $D_{\min}$
of artifact recurrence.

Let $\hat{d}[k]$ be the estimate of the disturbance signal.
We then define
\begin{equation}
    \hat{x}[k] = z[k] - \hat{d}[k] = h(z[k]; \vec{p})
\end{equation}
to be the estimated signal.
Here $h(\cdot)$ denotes the filtering function applied to the measurement $z[k]$, given the
cleaning parameters $\vec{p}$.
The parameters $\vec{p}$ depend on the specific algorithm used.

\subsection{Formal Process Description}

\begin{figure}
    \centering
    \includegraphics[width=\linewidth]{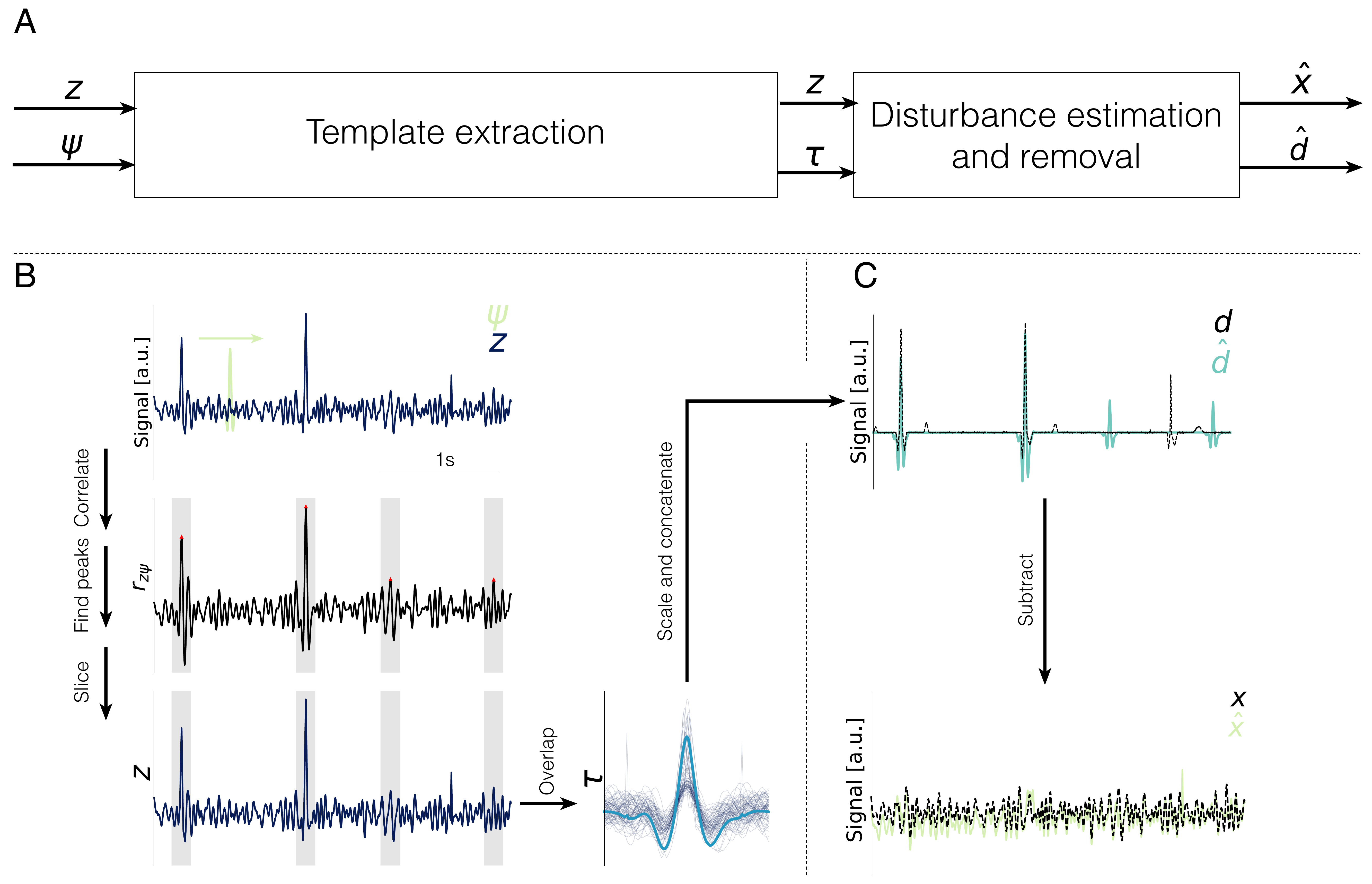}
    \caption{Algorithm overview.
    \textbf{A} The algorithm can be seen as a two-step solution.
    First, the template $\tau$ is extracted from the measurement $z$, based on the cross-correlation peaks $r_{z\psi}$ of $z$ and the search template $\psi$.
    Second, the template is concatenated and scaled to generate the disturbance estimate signal $\hat{d}$ which is subtracted from $z$ to yield the estimate $\hat{x}$.
    \textbf{B} The template extraction step involves the correlation of the search template with the measurement, then the peaks (red marks) are extracted, and the measurement is
    sliced around the cross-correlation peaks (gray areas). The slices are overlapped and averaged.
    \textbf{C} The template is scaled in correspondence of the cross-correlation peaks to reflect the local artifact strength. The scaled templates are then concatenated to form
    $\hat{d}$ and this is then subtracted from the measurement to obtain $\hat{x}$.
    }
    \label{fig:algo-overview}
\end{figure}

We now proceed with outlining the disturbance estimation and removal method.
Let $z[k]$ be the noisy measurement of $x[k]$, corrupted by additive disturbance $d[k]$ and noise $n[k]$.
Let $\psi[k]$ be a reasonable approximation of the dominant artifact introduced by the disturbance $d[k]$, i.e. assume $d[k]$ to be reasonably approximated by the concatenation (with ``pauses'') of scaled versions of $\psi[k]$:
\begin{equation}
    d[k] \approx \sum_m c_m\cdot\psi[k-k_m].
\end{equation}
Let $D_{\min}$ be the minimum number of samples in-between dominant artifacts in $d[k]$.
Then, the proposed algorithm operates by performing the following steps:
\begin{enumerate}
    \item[\textit{Optional}] The measurement $z[k]$ is band-pass filtered to only contain frequencies of interest to the analysis, e.g. removal of DC component/slow drifts with high-pass filtering, and/or removal of high frequency noise using a low-pass filter.
    \item Compute the cross-correlation $r_{\psi z}[k]$ of $z[k]$ and $\psi[k]$.
    \item Find the indices of the cross-correlation peaks $k_{pr,i}$ such that
    \begin{equation}
        r_{\psi z}[k_{pr,i}] > r_{\min} \wedge D_{\min} < k_{pr,i+1} -  k_{pr,i},
    \end{equation}
    with $r_{\min} > 0$ being the correlation peak threshold, i.e. we look for correlation peaks above a certain value $r_{\min}$ and having a given minimum distance $D_{\min}$ in-between each other.
    \item Extract the artifact template $\tau[k]$, by averaging the measurement signal windows around the cross-correlation peaks, i.e. average all slices $z[k_{pr,i}-K/2:k_{pr,i}+K/2]$, with $K$ being the template size.
    Shift, and scale $\tau$ to start and end at 0, and to have unit energy.
    \item Build the artifact signal $\hat{d}[k]$ by concatenating scaled version of the template at the locations of the cross-correlation peaks:
    \begin{equation}
        \hat{d}[k] = \sum_m r_{\tau z}[k_{pr,m}]\cdot\tau[k-k_{pr,m}].
    \end{equation}
    \item Estimate the clean signal as the difference of the measurement and the artifact signal (disturbance estimate):
    \begin{equation}
        \hat{x}[k] = z[k] - \hat{d}[k].
    \end{equation}
\end{enumerate}
Figure~\ref{fig:algo-overview} illustrates this procedure conceptually on sample data.
The presented cleaning algorithm $h(\cdot;\vec{p})$ draws inspiration from
the wavelet~\cite{mallat1999wavelet} and shapelet~\cite{ye2009time} analyses, hence
we refer to it as the wavelet template (WT) method from hereon.
The algorithm relies on the shape of the recurring disturbance pulse to ``resemble''
a (real-valued, sampled) wavelet $\psi[k]$.
An open-source implementation is made available as the \texttt{magnes-artifact-removal} package
at \url{https://github.com/magnesag/artrem}.

\subsection{Example Application}

We focus on the application of the presented formalization on a practical case:
the removal of heart artifacts from local-field potential (LFP) signals, corrupted
by additive pink-noise.
The rationale behind this choice is of historic origin:
the work presented herein is the formalization of work done on the analysis of
LFPs in subjects with (adaptive) deep-brain stimulation (DBS) implants.
In this domain, the removal of ECG artifacts can be essential for meaningful
data analysis results~\cite{neumann2021sensitivity}.
This use-case falls into the outlined framework, as the analysis of LFPs can be focused
on the $\beta$-waves spectrum of LFPs (frequency range 10--30\,Hz), the heart-beat can
induce artifacts in the LFP recording~\cite{neumann2021sensitivity,stam2023comparison},
dominating in the frequency range 5--15\,Hz~\cite{pan1985real}, and only one measurement channel can be
available in some cases~\cite{alberts2008are}.
It is important to mention that this work focuses on the identification of ECG as
artifact in LFP signals, and is not intended as a novel method for ECG analysis itself,
although some concepts may be transferred to this other domain.

\subsubsection{Ground-Truth Synthesis}
Given a unit-variance, ground-truth signal $x[k]$,
a comparable\footnote{Meaning that the peaks of $d[k]$ have amplitude in the
same range as the standard deviation of $x[k]$, which for zero-mean signals
corresponds to the root-mean-square of the signal.}
disturbance signal $d[k]$,
and some, zero mean, unit-variance generated noise $n[k]$, $\var(n) = 1$,
we define the measurement $z[k]$ as in equation (\ref{eq:measurement}).
The gain parameters $\gamma, \sigma > 0$ can be interpreted as
the z-score-gain for the respective addend.
Also, note that no gain for the ground-truth signal is used, as the relative
gains between $x$, $d$, and $n$ are relevant for the following analysis, and
not their absolute values.
This means that $\sigma$ can be linked to the noisy signal $x+\sigma\cdot n$
signal-to-noise ratio (SNR) as
\begin{equation}
    \snr = \frac{\var(x)}{\var(n)}=\frac{1}{\sigma^2}.
\end{equation}

One can come up with various methods for generating ground-truth signals.
We use the superposition of frequency-domain windowed white noise:
white noise time series are generated, transformed into frequency domain,
where a Gaussian window at a given center frequency and width is applied,
the signal is then transformed back into time-domain.
This method is aligned with the analysis performed in~\cite{donoghue2020parameterizing}
and the source code is made freely available as open-source Python module
\texttt{magnes-signal-generation-utility} available at \url{https://github.com/magnesag/siggen}.

\subsubsection{ECG as Disturbance}
The signal synthesis module is set up to generate disturbance signals from ECG
signals.
The signals are generated as a concatenation of plausible ECG pulses
constructed from pseudo-randomly sampled parameters.
Each pulse is generated to feature reasonable temporal properties and peak
amplitudes, for more information see the \texttt{cardio} submodule of \texttt{magnes-signal-generation-utility}
at \url{https://github.com/magnesag/siggen}.

\begin{figure}
    \centering
    \includegraphics[width=0.3\linewidth]{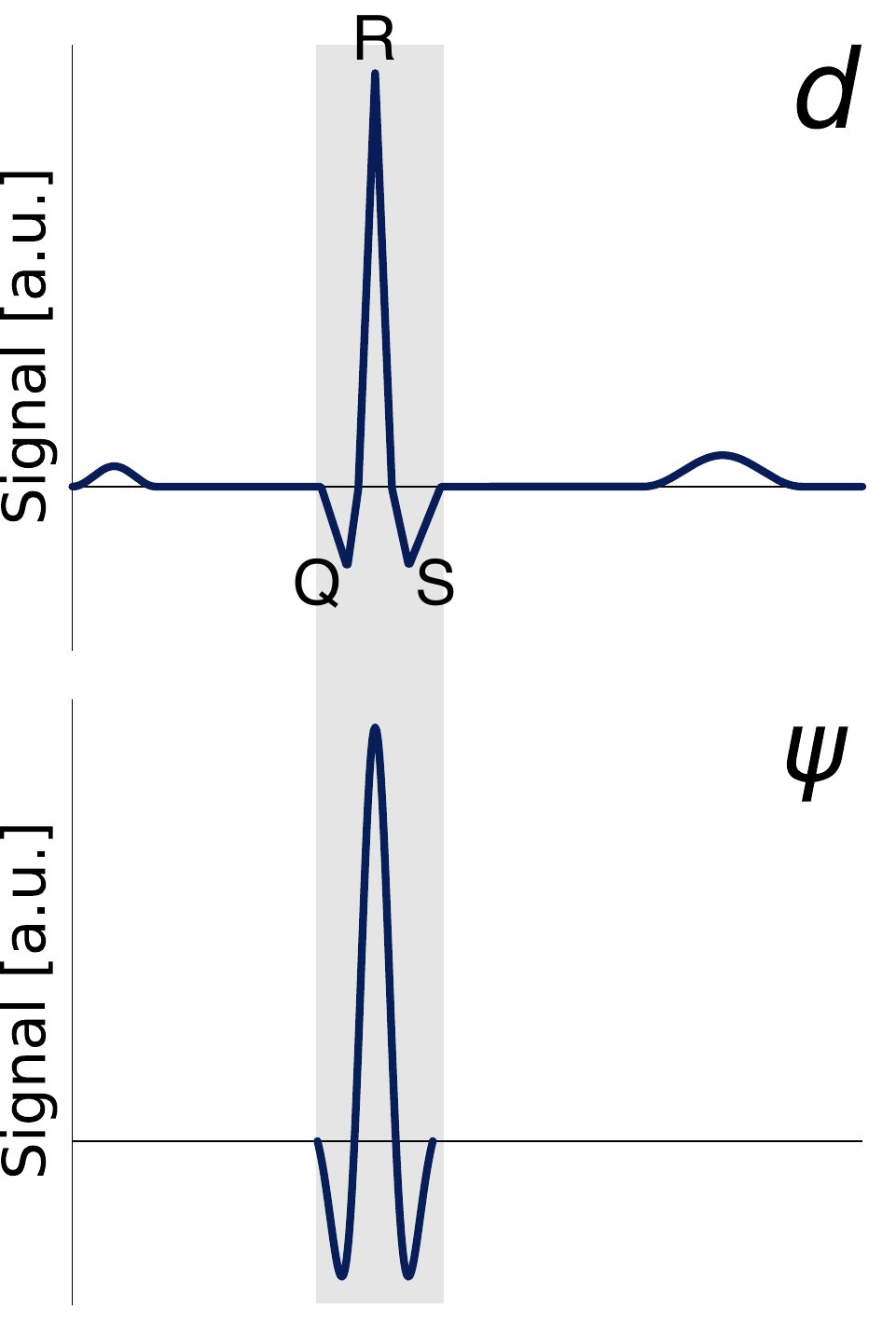}
    \caption{Example heart-beat signal $d$ and comparison to a Ricker wavelet instance $\psi$.
    The main origin of artifacts is the QRS complex (shaded gray area) with its characteristic shape.
    The resemblance between the two is apparent.}
    \label{fig:beat}
\end{figure}

In the investigated scenario, where heart-rate artifacts are to be removed from the
measurement $z[k]$, the disturbance signal $d[k]$ is assumed to be dominated by
the ECG QRS-complexes.
The QRS-complex has a typical shape as depicted in Figure~\ref{fig:beat},
and it has received extensive attention by cardiologists and engineers for its
detection and characterization~\cite{pan1985real,kohler2002principles,perez2016r}.
For the following analysis, it is noted that, typically, the QRS-complex has a
duration between 80 and 100 milliseconds.

\textbf{Picking a Search Template $\boldsymbol{\psi}$}
It is evident from Figure~\ref{fig:beat}, that the QRS-complex resembles a piecewise linear
version of the Ricker wavelet (also known as Mexican hat wavelet).
In continuous time terms, this wavelet is the negative second derivative of
the Gaussian bell:
\begin{equation}
    \psi(t) = -\frac{d^2 }{dt^2} \left( \exp\left(-a t^2\right) \right) = 2a \left(1 - 2at^2\right) \exp\left( -a t^2 \right) \label{eq:ricker}
\end{equation}
with $a = 1/2\sigma^2 > 0$, $\sigma > 0$.
We then define our artifact search-template $\psi[k]$, for $\sigma = 1 \iff a = 0.5$ to be
\begin{equation}
    \psi[k] = (1 - (k\cdot T_s)^2) \cdot \exp\left(-(k\cdot T_s)^2\right),\ k\cdot T_s = -3, ..., 3. \label{eq:search-template}
\end{equation}
That is, we generate the search-template from a unit-variance Gaussian and evaluate it
over the range [-3, 3], at the required sampling time $T_s$.
One can tune the search-template's number of samples $K$ for
a given sampling time $T_s$ to obtain a search-template of a given (real) duration $T$.
Specifically, one can use the relations
\begin{equation}
    T = (K+1) \cdot T_s \iff K = \frac{T}{T_s} + 1
\end{equation}
to determine the duration $T$ or the number of samples $N$, given the other and the sampling
time $T_s$.
It shall be noted that $\psi[k]$ as defined in (\ref{eq:search-template}) is shifted
and scaled after generation to meet the following constraints:
\begin{align*}
    & \psi[-K/2] = \psi[K/2] = 0,\\
    & T_s \cdot \sum_k \psi^2[k] = 1.
\end{align*}

\textbf{Frequency Domain Considerations for $\boldsymbol{\psi}$}
It can be shown\footnote{By evaluation of the Fourier transform $S(\omega)$ of equation (\ref{eq:ricker}) and
subsequently solving $dS/d\omega = 0$.} that for the chosen $\psi$, the peak frequency $\omega_p$ is at
\begin{equation}
    \omega_p = \pm \frac{6\sqrt{2}}{T},
\end{equation}
which, for a search-template duration of $T = 0.1\rm\,s$, which is a within the plausible
QRS-complex duration range,
this corresponds to a peak frequency of $\rm 84.85\,rad/s \approx 13.50\,Hz$.
While just outside the Pan-Tompkins band 5--12\,Hz~\cite{pan1985real},
this peak frequency falls within the range of interest of the
underlying ground-truth signal, for the LFP use-case, in particular the so-called
$\beta$-band, which spans the range 10--30\,Hz,
thus satisfying the problem requirement of $x$ and $d$ overlapping spectrally.
It shall be noted, that the $\beta$-band is of particular interest in the field of adaptive DBS,
see for example~\cite{thenaisie2021towards}.

\subsubsection{Machine-Learning Baseline}
We evaluate the performance of the proposed algorithm by comparing it with a machine-learning (ML) approach.
In particular, we train an autoencoder (AE) on synthetic data generated as previously
described and run both approaches on the same testing dataset, which is
purpose-generated and \textit{not} part of the training set of the AE.
This problem setting represents the ideal scenario for a data-driven approach -- the
training data is well representative of the test/deployment data, by definition.
Hence, the ML approach can provide a suitable benchmark against which the proposed
method can be compared to.

We use a narrow AE, with input-output window size of $M = 125$ samples
(0.5\,s at 250\,Hz sampling) and a single bottleneck, hidden layer of size 62
(compression factor of 2, total number of parameters 15687), which is trained
to estimate the artifact signal from the measurement, i.e. the input of the AE
is $z[k]$ and the output is $\hat{d}[k]$.
Formally, the AE can be expressed as a mapping $\mathcal{A}$ such that:
\begin{equation*}
    \mathcal{A}: \mathbb{R}^M \rightarrow \mathbb{R}^M, z \mapsto d = \mathcal{A}(z; \theta).
\end{equation*}
The AE depends on a set of parameters $\theta$ over which optimization is performed during training:
\begin{align*}
    \theta^* = \arg\min_\theta L\left( d, \mathcal{A}(z; \theta) \right),
\end{align*}
with $L$ being the loss function.
So we use
\begin{equation*}
    \hat{d}[k - M/2:k + M/2] = \mathcal{A}\left(z[k - M/2:k + M/2]; \theta^*\right)
\end{equation*}
as benchmark estimate, against which to evaluate the proposed template method.

The AE is defined and trained using
\texttt{Scikit-learn}~\cite{pedregosa2011sklearn},
as fully connected, multi-layer perceptron with ReLU activation.
Specifically, the AE is trained with mean square loss, Adam optimizer,
adaptive learning rate (initial rate 0.001) using early stopping (validation fraction 0.1),
auto batch size and at most 200 epochs~\cite{pedregosa2011sklearn},
feeding synthetic data with 1--10 random peak
frequencies $\vec{f}$ and widths $\vec{w}$, for noise and disturbance gains
$\sigma \in \{0.0, 0.4, 0.8, 0.8, 1.2, 1.6, 2.0\}$
$\gamma \in \{8.0, 6.6, 5.2, 3.8, 2.4, 1.0\}$ respectively.
For each $(\vec{f}, \vec{w}, \gamma, \sigma)$ tuple,
a 50-beats (artifacts) random time series\footnote{At a maximum heart-rate of 180\,BPM,
this corresponds to 16.7 seconds of data.}
sampled at $250\,Hz$ is generated,
out of which 50\% overlapping slices of $T=0.5\rm\,s$ are extracted to build the training dataset,
resulting in approximately 83k training samples.
It shall be noted that this definition of signals, i.e. starting from the contained number of artifacts,
ensures that the data is balanced in terms of disturbances, which is a desirable trait in the
training data.
Testing of the trained AE is performed on \textit{ad hoc} generated data during the performance evaluation and
comparison to the WT estimator.
It shall be noted that, the training data is fed starting from high to low
$\gamma/\sigma$ ratio to
incentivize learning of artifact recognition first, and noise rejection later
(targeted local minimum biasing).
This is achieved as follows.
The peak frequencies and peak widths $\vec{f}$ and $\vec{w}$ are varied jointly (only one width is considered for each peak frequency) and thus can be seen as a single hyperparameter $\phi=(\vec{f}, \vec{w})$.
In terms of signal generation, we thus have the hyperparameters $(\phi, \gamma, \sigma)$.
The training data is shuffled along the $\phi$ hyperparameter in order to
remove temporal and signal frequency dependency, but to keep order in
disturbance gain ($\gamma$, learning is started with high gains, which are easier
to learn) and noise gain ($\sigma$, learning begins without any additional noise,
and noise is gradually added as the model learns - as a sort of parameter fine-tuning
for increased robustness against noise as training advances).

\subsection{Performance Evaluation}

Given the synthetic data generation approach, we can evaluate the performance of the artifact
removal strategy proposed herein precisely.
We are in the position of being able to compare the estimate $\hat{x}[k]$ to the ground truth
$x[k]$ as well as the estimate of the disturbance $\hat{d}[k]$ to the ground truth $d[k]$.

We evaluate the performance of the method both in time- and frequency-domain.
Given the (partially) overlapping spectra for the disturbance and ground-truth, it
is important to be able to estimate the performance in frequency-domain.
As time-domain performance metric, we report the root mean squared error (RMSE)
between the estimate $\hat{x}[k]$ and the ground truth $x[k]$:
\begin{equation}
    \rmse(x, \hat{x}) = \sqrt{\expected_k\left( (\hat{x}[k] - x[k])^2 \right)},
\end{equation}
with $\expected()$ being the expected value operator, which reduces to the sample mean
for the numerical implementation.
We also evaluate the RMSE for $\hat{d}[k]$ with respect to $\gamma\cdot d[k]$.
In frequency-domain, we use the root mean square log error (RMSLE) as defined in~\cite{chen2021removal}
\begin{equation}
    \rmsle(x, \hat{x}) = \sqrt{\expected_f\left( 10\cdot\log_{10}\left( \frac{P_{x}(f)}{P_{\hat{x}}(f)} \right) \right)},
\end{equation}
where $P_x$ and $P_{\hat{x}}$ are the estimated power-spectra of $x$ and $\hat{x}$, i.e. the squared
magnitude of the Fourier transforms of $x$ and $\hat{x}$.
We also evaluate the RMSLE for $\hat{d}$ with respect to $\gamma \cdot d[k]$
The RMSLE can be interpreted as the RMSE of the log-power over the entire spectrum
(integration along the frequency-dimension).
For the estimation of the power-spectra, Welch's method~\cite{welch1967use} is used (\texttt{scipy.signal.welch()}).

When considering the underlying signal $x$ and its estimate $\hat{x}$,
the approach illustrated herein also enables us to evaluate the performance of the
cleaning with respect to the baseline of not-applying any filtering.
As described in~\cite{neumann2021sensitivity}, there are scenarios for which the
same type of measurement can or cannot be affected significantly by a disturbance.
Given the desire to automate the data analysis pipeline in order to be able to
perform large-scale studies, it is therefore important to gauge how the WT
(and AE) perform compared to no filtering at all.
While this may sound trivial at first, it shall be noticed that in a scarcely affected
measurement, it could be that cleaning introduces distortions instead of removing
them, as it is wrongly assumed that disturbances are affecting the measurement.
The performance of the presented algorithm in this sense, with respect to its
mixing parameters $\gamma$ and $\sigma$ is evaluated by comparing the RMSE and RMSLE
of the clean signal $\hat{x}[k]$ to the RMSE and RMSLE of the raw measurement $z[k]$,
both with respect to the ground-truth $x[k]$.
We denote the difference of performance metric with respect to the baseline with a $\Delta$, i.e.
$\Delta\rmse = \rmse(x, \hat{x}) - \rmse(x, z)$ and $\Delta\rmsle = \rmsle(x,\hat{x}) - \rmsle(x,z)$.
Hence, negative delta-metrics denote an improvement of the estimate with respect to the raw measurement,
which should be the objective of any algorithm of this sort.

The other performance metrics we use to evaluate the filtering strategy, are the
fraction of disturbance signal peaks detected as true-positives (correctly
removed artifacts) and false-positives (misfirings).
We define the number of positives to be the number of peaks in the artifact
estimate $\hat{d}$, which we denote as $\hat{N}$, while the ground truth
positives is the number of peaks in $d$, which we denote as $N$.
The true positives $\hat{N}_{TP}$ is then defined as the number of matching peaks
-- a matching peak is defined as a peak at the same index being detected in both
$d$ and $\hat{d}$.
The false positives $\hat{N}_{FP}$ are then defined to be the difference between
all detected peaks and the true positives $\hat{N}_{FP} = \hat{N} - \hat{N}_{TP}$.
The true positive rate (TPR) and the false positive rate (FPR) are then evaluated
as the fraction of true and false positives over the number of true positives and total positives
$\tpr = \hat{N}_{TP}/N$ and $\fpr = \hat{N}_{FP}/\hat{N}$.
This adaptation of the TPS and FPR is necessary for the context,
as the algorithm does not provide actual negative estimates,
but only absence of positives (detections).

\subsection{Parametric Sweep}

The proposed synthetic data generation approach gives us full control and knowledge
over the signal components.
At the same time though, two mixing parameters $\gamma$ and $\sigma$ have been introduced.
We assess the influence of the choice of the values of these parameters evaluating
the cleaning algorithm over a grid of $(\gamma, \sigma) \in \Gamma \times \Sigma$ values, while keeping the
other values constant.
$\Gamma$ and $\Sigma$ are the sets of all $\gamma$ and all $\sigma$ values considered in the search.
It is expected that larger mixing gains cause larger deviations from the ground-truth,
but that overall the cleaning algorithm yields lower deviations from the ground-truth
compared to the baseline.
It is to be expected that larger values of $\gamma$ ease the artifact detection, while
larger values of $\sigma$ lower the accuracy.

In order to take into account the random nature of the synthetic data, we evaluate the full
parameter grid $(\gamma, \sigma) \in \Gamma \times \Sigma$ for $N$ randomly generated datasets.
For each dataset, we randomly set the number and location of the peak frequencies, as well as
the width of the peaks.

  \section{Results}\label{sec:results}

\begin{figure}
    \centering
    \includegraphics[width=0.75\linewidth]{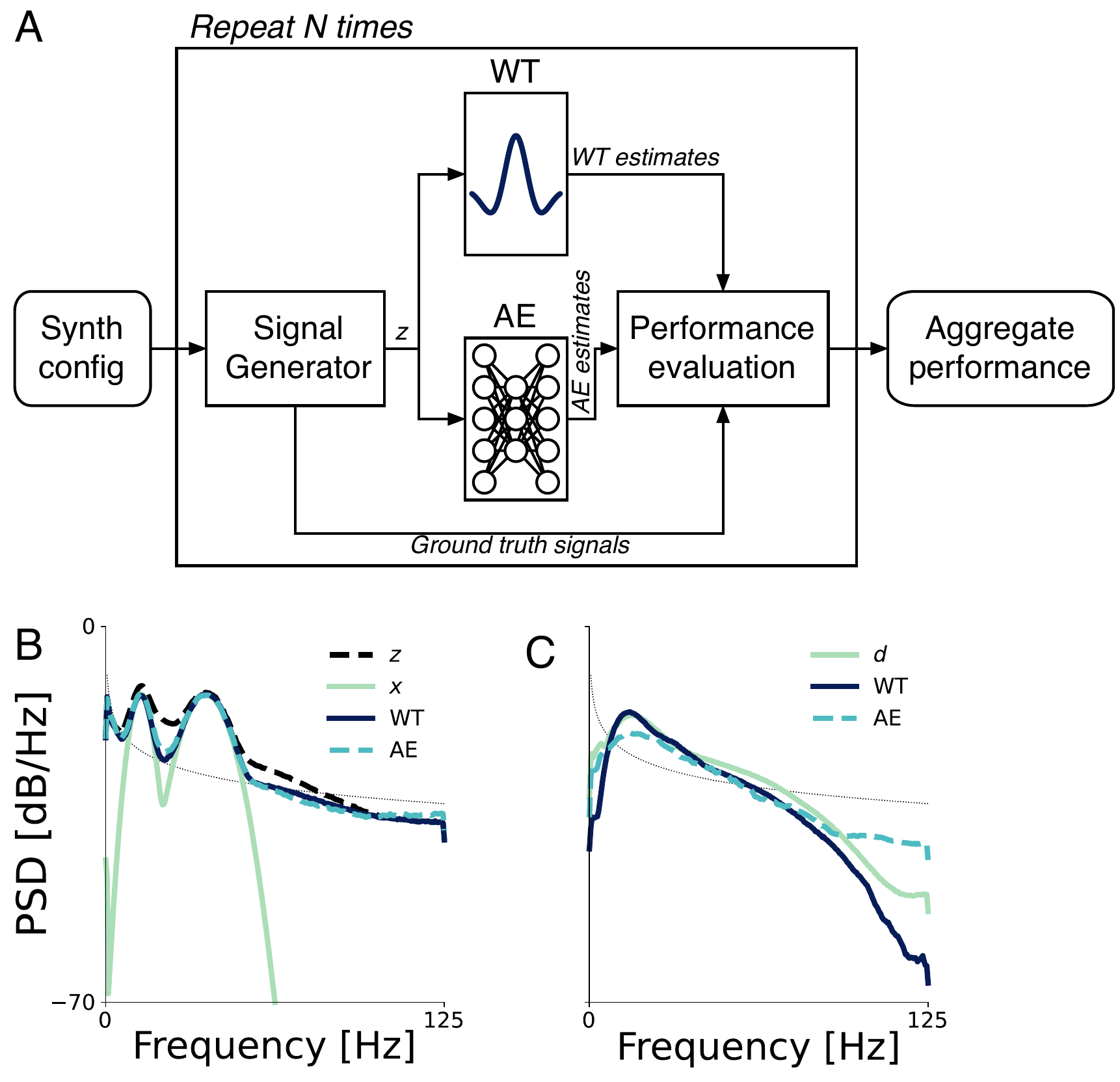}
    \caption{Frequency-domain results on synthetic data.
    The algorithm is able to detect and remove the artifacts from the measurement $z$,
    even in the presence of (strong) noise.
    The median spectra over $N=100$ evaluations of simulated signals at constant configuration
    parameters are shown.}
    \label{fig:fd-synth-comp}
\end{figure}

\textbf{Single Synthesis Configuration}
We generate 100 synthetic signals with the following configuration:
$\vec{f} = (13.0, 37.0)$ Hz,
$\vec{w} = (2.0, 4.2)$ Hz,
$\sigma = 0.5$,
$\gamma = 8.0$.
The rationale for this choice is as follows:
the selected frequency parameters $\vec{f}$ and $\vec{w}$ create the desired spectral
overlap with $d$ (peak at around 13.5\,Hz), while the chosen values for $\sigma$ and $\gamma$
ensure the presence of significant noise ($\snr = 4$), but also clear artifacts.
We then run the WT and AE on all datasets and evaluate the performance in
terms of RMSE, RMSLE, TPR and FPR.
A frequency-domain visualization of the results is shown
in Figure~\ref{fig:fd-synth-comp}.
The comparative results are reported in Table~\ref{tab:singe-config-results}.

It shall be noted that the RMSLE is evaluated over the entire frequency range, and is thus
dominated by the difference due to noise, i.e. even a small change in RMSLE is an
important step forward.

By comparing the proposed WT method to the AE approach, the proposed method
represents a valid alternative to the AE.
In particular, the proposed method yields a $1-0.59/(0.59+0.29) = 33\%$ median
improvement in signal estimate RMSE compared to the
$1-0.63/(0.63+0.26) = 29\%$ median improvement brought by the AE.
One can clearly see in Figure~\ref{fig:fd-synth-comp} how the both methods
closely approximate the disturbance spectrum around its peak, i.e. where its effect is the greatest.

The AE tends to show smaller differences between its median $m$ and its 95th percentile $P_{95}$ scores,
hinting towards a greater consistency with respect to the WT (Table~\ref{tab:singe-config-results}).
Nevertheless, for the particular set of parameters, the WT actually performs better than the AE on average
for almost all metrics, exception be made for the TPR.

\begin{table}
    \centering
    \caption{Artifact estimation and removal performance for $N=100$ simulations carried out with the same configuration.
    Values are reported as median $m$ and 95th percentile $P_{95}$.}
    \label{tab:singe-config-results}
    \begin{tabular}{lrrrrrrrr}
        \toprule
        & \multicolumn{4}{c}{WT} & \multicolumn{4}{c}{AE} \\
        \cmidrule(r){2-5}
        \cmidrule(l){6-9}
        & \multicolumn{2}{c}{$x$} & \multicolumn{2}{c}{$d$} & \multicolumn{2}{c}{$x$} & \multicolumn{2}{c}{$d$}\\
        \cmidrule(r){2-3}
        \cmidrule(lr){4-5}
        \cmidrule(lr){6-7}
        \cmidrule(l){8-9}
         & $m$ & $P_{95}$ & $m$ & $P_{95}$ & $m$ & $P_{95}$ & $m$ & $P_{95}$ \\
        \midrule
        RMSE & $\boldsymbol{0.59}$ & $0.83$ & $\boldsymbol{0.35}$ & $\boldsymbol{0.38}$ & $0.63$ & $\boldsymbol{0.64}$ & $0.39$ & $0.41$\\
        \midrule
        $\Delta$RMSE & $\boldsymbol{-0.29}$ & $-0.10$ & \multicolumn{2}{c}{--} & $-0.26$ & $\boldsymbol{-0.24}$ & \multicolumn{2}{c}{--}\\
        \midrule
        RMSLE & $\boldsymbol{70.16}$ & $\boldsymbol{70.90}$ & $5.78$ & $7.69$ & $70.27$ & $70.94$ & $\boldsymbol{4.33}$ & $\boldsymbol{4.80}$ \\
        \midrule
        $\Delta$RMSLE & $\boldsymbol{-0.70}$ & $-0.49$ & \multicolumn{2}{c}{--} & $-0.66$ & $\boldsymbol{-0.55}$ & \multicolumn{2}{c}{--} \\
        \midrule
        TPR & \multicolumn{2}{c}{--} & $0.76$ & $0.86$ & \multicolumn{2}{c}{--} & $\boldsymbol{0.82}$ & $\boldsymbol{0.92}$ \\
        \midrule
        FPR & \multicolumn{2}{c}{--} & $\boldsymbol{0.33}$ & $\boldsymbol{0.43}$ & \multicolumn{2}{c}{--} & $0.44$ & $0.54$ \\
        \bottomrule
    \end{tabular}
\end{table}


\begin{figure}
    \centering
    \includegraphics[width=\linewidth]{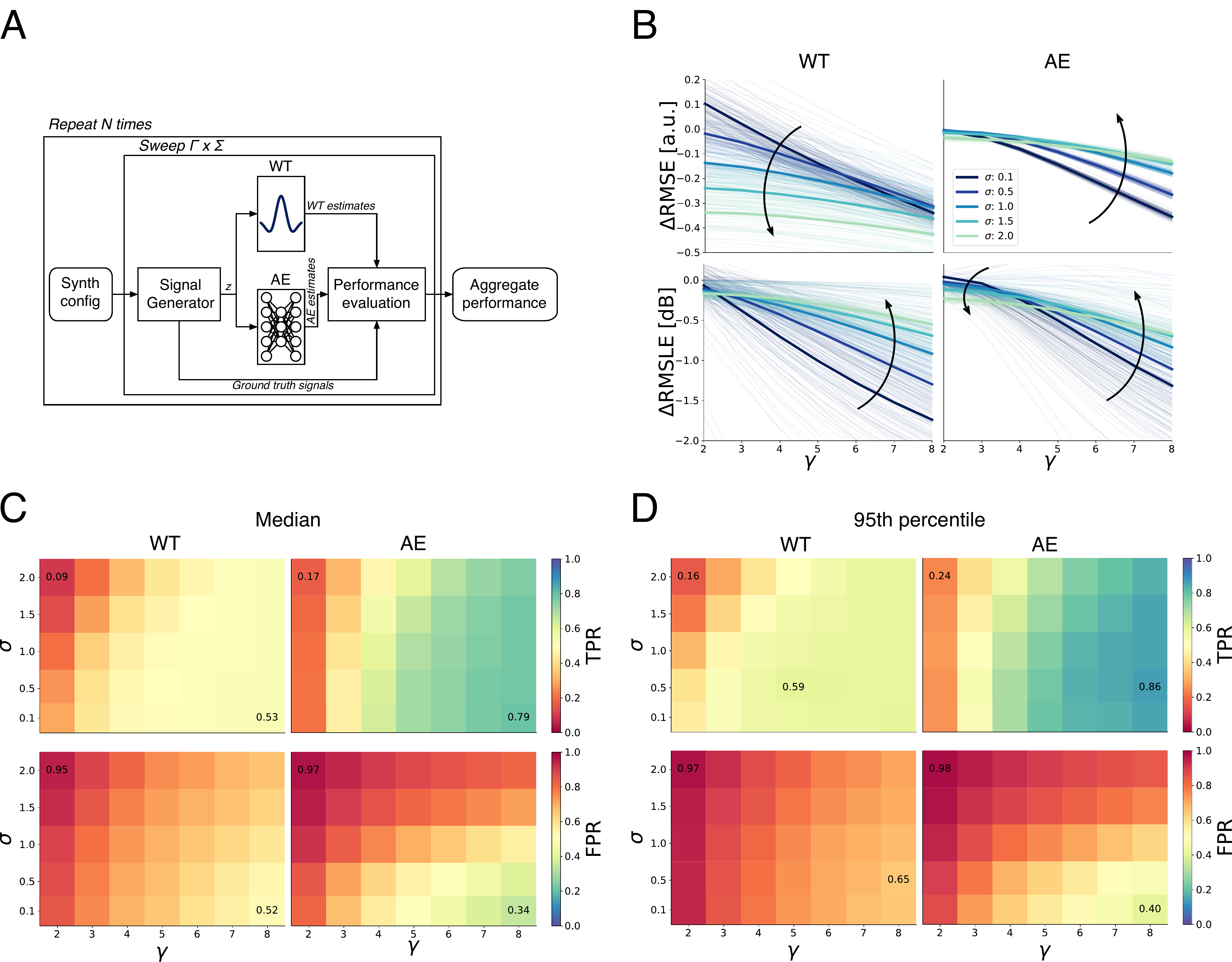}
    \caption{Parametric sweep results.
    \textbf{A} Sweep process graphical visualization: notice that the signal synthesis configurations are generated $N$ times, creating $N$ different $(\vec{f}, \vec{w}, d, n)$ instances, for each of which the entire $\Gamma \times \Sigma$ space is evaluated.
    \textbf{B} $\Delta\rmse$ and $\Delta\rmsle$ for the true signal $x$ for 100 synthetic signals and parametric sweep on $\Gamma \times \Sigma$. This plot show higher consistency for the AE, but also how the WT can actually outperform the AE under certain conditions and metrics, e.g. in terms of RMSE for larger values of $\sigma$.
    The thin lines show the results for each experiment, while the thick lines represent the median.
    The arrows indicate increasing $\sigma$ directions and are meant to ease reading of the plot.
    \textbf{C} and
    \textbf{D} show the TPR and FPR for the two methods as median values and 95th percentile values respectively. The minimum and the maximum value are reported for each metric and method.
    These plots show how the AE performs better than the WT in terms of artifact detection in a pure on-off sense.}
    \label{fig:sweep-res}
\end{figure}

\textbf{Parametric Sweep}
We ran the parametric sweep for $\Sigma = \{0.1, 0.5,1.0,1.5, 2.0\}$ and $\Gamma = \{ 2.0,3.0,4.0, \dots, 8.0 \}$.
$N = 100$ experiments have been run and the results aggregated.
In each experiment, a random frequency $\vec{f}$ and width $\vec{w}$ sets were generated to define the
ground truth signal $x$, alongside random base disturbance $d$ and noise $n$.
The parameter grid $\Gamma \times \Sigma$ was then swept (in a grid-search-like manner) to generate
the measurement signals $z = x + \gamma \cdot d + \sigma \cdot n$.
Hence, for each $(\gamma, \sigma) \in \Gamma \times \Sigma$, $N$ measurements and performance evaluations have
been performed.
The performance scores across the experiments are visualized in Figure~\ref{fig:sweep-res}.
The sweep confirms the fact that the AE shows greater consistency when compared to the WT.
Nonetheless, in terms of metrics and individual runs, it can be seen (Figure~\ref{fig:sweep-res}) that
the proposed WT method can yield better results than the AE.

  \section{Discussion}

The presented algorithm has shown to effectively remove artifacts from
disturbed signals.
In particular its performance has been proven to be satisfactory for
cases in which the artifacts are dominant.
As expected, the both methods, wavelet template (WT) and autoencoder (AE),
perform best when the disturbance dominates
the measurement, i.e. $\gamma > 1 \wedge \gamma > \sigma$.

The WT method has been proven to yield results that can even be better than
what can be obtained using an AE.
Despite the AE being used in an idealized context, the WT outperforms the AE
in some regards.
The WT RMSE gains compared to the baseline are greater for the WT than the AE
and the FPR can be better for the WT in some cases.
Overall, the AE does yield improved consistency over the proposed method,
but is must be remembered that the AE is deployed in an ideal scenario:
the data it is being tested with is generated with the same methods as the
data it was trained on, hindering the possibility of generalizability.
This of course is rarely if ever the case in real-world applications.
Therefore, being the WT able to yield such relatively good results in
comparison to the AE is quite remarkable.

While the biggest limitation for the AE approach is its generalizability, i.e. the
need for good training data to achieve useful results and applications in real-world
scenarios, it has to be remembered that the proposed algorithm also features some
\textit{a priori} knowledge of the disturbance.
Though, the need for an initial shape for the disturbance is regarded as a minor hurdle
-- a reasonable guess has been proven to yield satisfactory results, and
it shall be noted that the actual template is extracted by the algorithm from the
measurement itself, thus partially decoupling the choice from the performance.

The main limitation of the algorithm, remains that artifacts will
always be removed, regardless of their actual presence.
When artifacts are comparatively small, i.e. $\gamma \rightarrow 1^+$, the
methods tend to perform worse, albeit potentially still yielding improvements
over the baseline at least in spectral terms (see Figure~\ref{fig:fd-synth-comp}).
This means that one needs to first evaluate whether there is the need
for artifact removal, \textit{before} using its results.
Similar conclusions have been drawn in previous works~\cite{stam2023comparison},
with the choice on whether to apply artifact removal is left to human
judgement on a case-by-case basis.
Future efforts should therefore consider focusing on the automatic decision-making
in this sense.
This would bring the benefit of easing the data-analysis process enabling
larger studies to be carried out.
Also, the clinical use could be accelerated, as fully automated pipelines
could be implemented enabling clinicians to only focus on the interpretation
of measurements and not worry about the data acquisition.
  \section{Conclusions}\label{sec:conclusions}

A formalized method for the removal of artifact-like disturbances from signals,
based on a data-augmented wavelet template (WT) has been introduced.
The WT method has been compared to the \textit{de facto} standard for the
task under consideration, i.e. the autoencoder.
The two methods have been compared extensively, including a parametric sweep
on the measurement parameters.
It has been shown that the proposed WT method can be a valid alternative to an
AE.
The WT also features important advantages compared to the AE:
1) the WT does not suffer from generalization uncertainty; and
2) the WT can easily be adapted to various scenarios without expensive retraining.
The WT does require an initial guess on the shape of the artifact,
but this is in no way comparable to designing and (re)training of an AE for a specific
task.

The proposed method, as well as the methods to generate the signals to
evaluate and test algorithms on have been released into the public domain.
The source code is freely available.
The authors hope that this will trigger further developments and investigations
in the field.

  \setlength{\parindent}{0pt}

  \section*{Author Contributions (CRediT)}
  AS: conceptualization, software, formal analysis, investigation, methodology, visualization, writing -- original draft;
  HM: conceptualization, software, formal analysis, methodology, writing -- review \& editing;
  GC: funding acquisition, project administration, writing -- review \& editing;
  HT: writing -- review \& editing.
  All authors have read and agreed to the submitted version of the manuscript.


  \section*{Acknowledgements}
  The authors would like to thank
  Carlo Mangiante from Magnes AG, Switzerland,
  for the valuable comments and inputs on the manuscript
  and Olgac Ergeneman from Magnes AG, Switzerland,
  for the additional administrative coordination.

  \section*{Data and Code Availability}
  All data was synthetically generated using the open-source code found at: \url{https://github.com/magnesag/siggen}.
  The analysis SW is available at: \url{https://github.com/magnesag/artrem}.

  \section*{Conflicts of Interest}
  The authors declare no conflicts of interest.

  \bibliographystyle{unsrt}
  \bibliography{bibliography}

@article{alberts2008are,
  title     = {Are two leads always better than one: an emerging case for unilateral subthalamic deep brain stimulation in {P}arkinson's disease},
  author    = {Alberts, Jay L and Hass, Christopher J and Vitek, Jerrold L and Okun, Michael S},
  journal   = {Experimental neurology},
  volume    = {214},
  number    = {1},
  pages     = {1--5},
  year      = {2008},
  publisher = {Elsevier}
}

@inproceedings{almazrouei2019using,
  title     = {Using autoencoders for radio signal denoising},
  author    = {Almazrouei, Ebtesam and Gianini, Gabriele and Mio, Corrado and Almoosa, Nawaf and Damiani, Ernesto},
  booktitle = {Proceedings of the 15th ACM International Symposium on QoS and Security for Wireless and Mobile Networks},
  pages     = {11--17},
  year      = {2019}
}

@article{brunelli1997template,
  title     = {Template matching: Matched spatial filters and beyond},
  author    = {Brunelli, Roberto and Poggiot, T},
  journal   = {Pattern recognition},
  volume    = {30},
  number    = {5},
  pages     = {751--768},
  year      = {1997},
  publisher = {Elsevier}
}

@article{chen2021removal,
  author   = {Chen, Yue  and Ma, Bozhi  and Hao, Hongwei  and Li, Luming },
  title    = {Removal of Electrocardiogram Artifacts From Local Field Potentials Recorded by Sensing-Enabled Neurostimulator},
  journal  = {Frontiers in Neuroscience},
  volume   = {Volume 15 - 2021},
  year     = {2021},
  url      = {https://www.frontiersin.org/journals/neuroscience/articles/10.3389/fnins.2021.637274},
  doi      = {10.3389/fnins.2021.637274},
  issn     = {1662-453X}
}

@article{cheng2021improved,
  title     = {Improved autoencoder for unsupervised anomaly detection},
  author    = {Cheng, Zhen and Wang, Siwei and Zhang, Pei and Wang, Siqi and Liu, Xinwang and Zhu, En},
  journal   = {International Journal of Intelligent Systems},
  volume    = {36},
  number    = {12},
  pages     = {7103--7125},
  year      = {2021},
  publisher = {Wiley Online Library}
}

@article{donoghue2020parameterizing,
  title     = {Parameterizing neural power spectra into periodic and aperiodic components},
  author    = {Donoghue, Thomas and Haller, Matar and Peterson, Erik J and Varma, Paroma and Sebastian, Priyadarshini and Gao, Richard and Noto, Torben and Lara, Antonio H and Wallis, Joni D and Knight, Robert T and others},
  journal   = {Nature neuroscience},
  volume    = {23},
  number    = {12},
  pages     = {1655--1665},
  year      = {2020},
  publisher = {Nature Publishing Group US New York}
}

@article{hammer2022artifact,
  title     = {Artifact characterization and a multipurpose template-based offline removal solution for a sensing-enabled deep brain stimulation device},
  author    = {Hammer, Lauren H and Kochanski, Ryan B and Starr, Philip A and Little, Simon},
  journal   = {Stereotactic and functional neurosurgery},
  volume    = {100},
  number    = {3},
  pages     = {168--183},
  year      = {2022},
  publisher = {S. Karger AG}
}

@article{harris2020array,
  title     = {Array programming with {NumPy}},
  author    = {Charles R. Harris and K. Jarrod Millman and St{\'{e}}fan J.
               van der Walt and Ralf Gommers and Pauli Virtanen and David
               Cournapeau and Eric Wieser and Julian Taylor and Sebastian
               Berg and Nathaniel J. Smith and Robert Kern and Matti Picus
               and Stephan Hoyer and Marten H. van Kerkwijk and Matthew
               Brett and Allan Haldane and Jaime Fern{\'{a}}ndez del
               R{\'{i}}o and Mark Wiebe and Pearu Peterson and Pierre
               G{\'{e}}rard-Marchant and Kevin Sheppard and Tyler Reddy and
               Warren Weckesser and Hameer Abbasi and Christoph Gohlke and
               Travis E. Oliphant},
  year      = {2020},
  month     = sep,
  journal   = {Nature},
  volume    = {585},
  number    = {7825},
  pages     = {357--362},
  doi       = {10.1038/s41586-020-2649-2},
  publisher = {Springer Science and Business Media {LLC}},
  url       = {https://doi.org/10.1038/s41586-020-2649-2}
}

@article{hossain2022deep,
  title     = {A deep convolutional autoencoder for automatic motion artifact removal in electrodermal activity},
  author    = {Hossain, Md-Billal and Posada-Quintero, Hugo F and Chon, Ki H},
  journal   = {IEEE Transactions on Biomedical Engineering},
  volume    = {69},
  number    = {12},
  pages     = {3601--3611},
  year      = {2022},
  publisher = {IEEE}
}

@article{hunter2007matplotlib,
  author    = {Hunter, J. D.},
  title     = {{Matplotlib}: {A} {2D} graphics environment},
  journal   = {Computing in Science \& Engineering},
  volume    = {9},
  number    = {3},
  pages     = {90--95},
  publisher = {IEEE COMPUTER SOC},
  doi       = {10.1109/MCSE.2007.55},
  year      = 2007
}

@article{kohler2002principles,
  author   = {Kohler, B.-U. and Hennig, C. and Orglmeister, R.},
  journal  = {IEEE Engineering in Medicine and Biology Magazine},
  title    = {The principles of software {QRS} detection},
  year     = {2002},
  volume   = {21},
  number   = {1},
  pages    = {42-57},
  doi      = {10.1109/51.993193}
}

@article{kramer1992autoassociative,
  title     = {Autoassociative neural networks},
  author    = {Kramer, Mark A},
  journal   = {Computers \& chemical engineering},
  volume    = {16},
  number    = {4},
  pages     = {313--328},
  year      = {1992},
  publisher = {Elsevier}
}

@book{mallat1999wavelet,
  title     = {A wavelet tour of signal processing},
  author    = {Mallat, St{\'e}phane},
  year      = {1999},
  publisher = {Elsevier}
}

@article{neumann2021sensitivity,
  title     = {The sensitivity of {ECG} contamination to surgical implantation site in brain computer interfaces},
  author    = {Neumann, Wolf-Julian and Sorkhabi, Majid Memarian and Benjaber, Moaad and Feldmann, Lucia K and Saryyeva, Assel and Krauss, Joachim K and Contarino, Maria Fiorella and Sieger, Tomas and Jech, Robert and Tinkhauser, Gerd and others},
  journal   = {Brain Stimulation},
  volume    = {14},
  number    = {5},
  pages     = {1301--1306},
  year      = {2021},
  publisher = {Elsevier}
}

@article{pan1985real,
  title     = {A real-time {QRS} detection algorithm},
  author    = {Pan, Jiapu and Tompkins, Willis J},
  journal   = {IEEE transactions on biomedical engineering},
  volume    = {BME-32},
  number    = {3},
  pages     = {230--236},
  year      = {1985},
  publisher = {IEEE}
}

@article{pedregosa2011sklearn,
  title   = {Scikit-learn: Machine Learning in {P}ython},
  author  = {Pedregosa, F. and Varoquaux, G. and Gramfort, A. and Michel, V.
             and Thirion, B. and Grisel, O. and Blondel, M. and Prettenhofer, P.
             and Weiss, R. and Dubourg, V. and Vanderplas, J. and Passos, A. and
             Cournapeau, D. and Brucher, M. and Perrot, M. and Duchesnay, E.},
  journal = {Journal of Machine Learning Research},
  volume  = {12},
  pages   = {2825--2830},
  year    = {2011}
}

@article{perez2016r,
  title     = {R-peak time: {A}n electrocardiographic parameter with multiple clinical applications},
  author    = {P{\'e}rez-Riera, Andr{\'e}s Ricardo and de Abreu, Luiz Carlos and Barbosa-Barros, Raimundo and Nikus, Kjell C and Baranchuk, Adrian},
  journal   = {Annals of Noninvasive Electrocardiology},
  volume    = {21},
  number    = {1},
  pages     = {10--19},
  year      = {2016},
  publisher = {Wiley Online Library}
}

@article{saba2021unsupervised,
  title     = {Unsupervised {EEG} artifact detection and correction},
  author    = {Saba-Sadiya, Sari and Chantland, Eric and Alhanai, Tuka and Liu, Taosheng and Ghassemi, Mohammad M},
  journal   = {Frontiers in digital health},
  volume    = {2},
  pages     = {608920},
  year      = {2021},
  publisher = {Frontiers Media SA}
}

@article{stam2023comparison,
  title     = {A comparison of methods to suppress electrocardiographic artifacts in local field potential recordings},
  author    = {Stam, MJ and van Wijk, BCM and Sharma, P and Beudel, Martijn and Pi{\~n}a-Fuentes, DA and de Bie, RMA and Schuurman, PR and Neumann, W-J and Buijink, AWG},
  journal   = {Clinical Neurophysiology},
  volume    = {146},
  pages     = {147--161},
  year      = {2023},
  publisher = {Elsevier}
}

@article{sweeney2012artifact,
  title     = {Artifact removal in physiological signals -- {P}ractices and possibilities},
  author    = {Sweeney, Kevin T and Ward, Tom{\'a}s E and McLoone, Se{\'a}n F},
  journal   = {IEEE transactions on information technology in biomedicine},
  volume    = {16},
  number    = {3},
  pages     = {488--500},
  year      = {2012},
  publisher = {IEEE}
}

@article{thenaisie2021towards,
  title     = {Towards adaptive deep brain stimulation: clinical and technical notes on a novel commercial device for chronic brain sensing},
  author    = {Thenaisie, Yohann and Palmisano, Chiara and Canessa, Andrea and Keulen, Bart J and Capetian, Philipp and Jim{\'e}nez, Mayte Castro and Bally, Julien F and Manferlotti, Elena and Beccaria, Laura and Zutt, Rodi and others},
  journal   = {Journal of neural engineering},
  volume    = {18},
  number    = {4},
  pages     = {042002},
  year      = {2021},
  publisher = {IOP Publishing}
}

@article{virtanen2020scipy,
  author  = {Virtanen, Pauli and Gommers, Ralf and Oliphant, Travis E. and
             Haberland, Matt and Reddy, Tyler and Cournapeau, David and
             Burovski, Evgeni and Peterson, Pearu and Weckesser, Warren and
             Bright, Jonathan and {van der Walt}, St{\'e}fan J. and
             Brett, Matthew and Wilson, Joshua and Millman, K. Jarrod and
             Mayorov, Nikolay and Nelson, Andrew R. J. and Jones, Eric and
             Kern, Robert and Larson, Eric and Carey, C J and
             Polat, {\.I}lhan and Feng, Yu and Moore, Eric W. and
             {VanderPlas}, Jake and Laxalde, Denis and Perktold, Josef and
             Cimrman, Robert and Henriksen, Ian and Quintero, E. A. and
             Harris, Charles R. and Archibald, Anne M. and
             Ribeiro, Ant{\^o}nio H. and Pedregosa, Fabian and
             {van Mulbregt}, Paul and {SciPy 1.0 Contributors}},
  title   = {{{SciPy} 1.0: {F}undamental Algorithms for Scientific Computing in {P}ython}},
  journal = {Nature Methods},
  year    = {2020},
  volume  = {17},
  pages   = {261--272},
  adsurl  = {https://rdcu.be/b08Wh},
  doi     = {10.1038/s41592-019-0686-2}
}

@article{welch1967use,
  author   = {Welch, P.},
  journal  = {IEEE Transactions on Audio and Electroacoustics},
  title    = {The use of fast {F}ourier transform for the estimation of power spectra: {A} method based on time averaging over short, modified periodograms},
  year     = {1967},
  volume   = {15},
  number   = {2},
  pages    = {70-73},
  doi      = {10.1109/TAU.1967.1161901}
}

@article{xiong2015denoising,
  title     = {Denoising autoencoder for eletrocardiogram signal enhancement},
  author    = {Xiong, Peng and Wang, Hongrui and Liu, Ming and Liu, Xiuling},
  journal   = {Journal of Medical Imaging and Health Informatics},
  volume    = {5},
  number    = {8},
  pages     = {1804--1810},
  year      = {2015},
  publisher = {American Scientific Publishers}
}

@inproceedings{ye2009time,
  title     = {Time series shapelets: a new primitive for data mining},
  author    = {Ye, Lexiang and Keogh, Eamonn},
  booktitle = {Proceedings of the 15th ACM SIGKDD international conference on Knowledge discovery and data mining},
  pages     = {947--956},
  year      = {2009}
}

@inproceedings{zhou2017anomaly,
  title     = {Anomaly detection with robust deep autoencoders},
  author    = {Zhou, Chong and Paffenroth, Randy C},
  booktitle = {Proceedings of the 23rd ACM SIGKDD international conference on knowledge discovery and data mining},
  pages     = {665--674},
  year      = {2017}
}

\end{document}